\documentclass[
    aps,
superscriptaddress,
    prl, 
    twocolumn,
    letterpaper,
    10pt,
    floatfix,
    nofootinbib
]{revtex4-1}

\usepackage{enumerate,graphicx,amsmath,amssymb}
\usepackage{epstopdf}
\usepackage{siunitx}
\usepackage{easy-todo}
\usepackage{comment}
\usepackage[usenames,dvipsnames]{xcolor}
\usepackage[normalem]{ulem} 
\usepackage{url}
\usepackage{appendix}
\usepackage[colorlinks,
    citecolor=Fuchsia,
    linkcolor=TealBlue,
    urlcolor=Violet]{hyperref}
\usepackage[capitalize,nameinlink]{cleveref}

\def\beq{\begin{equation}}
\def\eeq{\end{equation}}

\newcommand{\delete}[1]{}

\newcommand{\be}{\begin{equation}}
\newcommand{\ee}{\end{equation}}

\begin{document}

\title{Searching for new physics with a levitated-sensor-based gravitational-wave detector}

\author{Nancy Aggarwal}
\affiliation{Center for Fundamental Physics, Department of Physics and Astronomy, Northwestern University, Evanston, Illinois 60208, USA}
\affiliation{Center for Interdisciplinary Exploration and Research in Astrophysics (CIERA), Department of Physics and Astronomy, Northwestern University, Evanston, Illinois 60208, USA}

\author{George P. Winstone}
\affiliation{Center for Fundamental Physics, Department of Physics and Astronomy, Northwestern University, Evanston, Illinois 60208, USA}

\author{Mae Teo}
\affiliation{Stanford Institute for Theoretical Physics, Stanford University, Stanford, California 94305, USA}

\author{Masha Baryakhtar}
\affiliation{Center for Cosmology and Particle Physics, Department of Physics, New York University, New York, NY 10003, USA}
\author{Shane L. Larson}
\affiliation{Center for Interdisciplinary Exploration and Research in Astrophysics (CIERA), Department of Physics and Astronomy, Northwestern University, Evanston, Illinois 60208, USA}

\author{Vicky Kalogera}
\affiliation{Center for Interdisciplinary Exploration and Research in Astrophysics (CIERA), Department of Physics and Astronomy, Northwestern University, Evanston, Illinois 60208, USA}

\author{Andrew A. Geraci}
\affiliation{Center for Fundamental Physics, Department of Physics and Astronomy, Northwestern University, Evanston, Illinois 60208, USA}
\affiliation{Center for Interdisciplinary Exploration and Research in Astrophysics (CIERA), Department of Physics and Astronomy, Northwestern University, Evanston, Illinois 60208, USA}

\begin{abstract}
The Levitated Sensor Detector (LSD) is a  compact resonant gravitational-wave (GW) detector based on optically trapped dielectric particles that is under construction. The LSD sensitivity has more favorable frequency scaling at high frequencies compared to laser interferometer detectors such as LIGO. We propose a method to substantially improve the sensitivity by optically levitating a multi-layered stack of dielectric discs. These stacks allow the use of a more massive levitated object while exhibiting minimal photon recoil heating due to light scattering. Over an order of magnitude of unexplored frequency space for GWs above 10 kHz is accessible with an instrument 10 to 100 meters in size. Particularly motivated sources in this frequency range are gravitationally bound states of QCD axions with decay constant near the grand unified theory (GUT) scale that form through black hole superradiance and annihilate to GWs. The LSD is also sensitive to GWs from binary coalescence of sub-solar-mass primordial black holes and as-yet unexplored new physics in the high-frequency GW window.
\end{abstract}
\maketitle

{\em Introduction---} 
The kilometer-scale gravitational-wave (GW) interferometers have just begun to view the universe in the domain of gravitational radiation, with remarkable sensitivity at frequencies ranging from 10s of Hz to a few kHz \cite{LIGOfirst2016}. Already, several exciting discoveries have resulted from these instruments, including the existence of binary black hole (BH) and neutron star systems \cite{LIGOVIRGO}. 
In this nascent field it is imperative to extend the GW search to other frequencies, just as x-ray- and radio-astronomy have done for the electromagnetic spectrum. Other promising experiments and techniques for probing the GW spectrum, including pulsar timing arrays \cite{PTA,SKA}, atomic clocks and interferometers \cite{GWclocks,MAGIS}, LISA \cite{LISA1,LISA2}, and DECIGO \cite{DECIGO}, all search at frequencies lower than the LIGO range. While at MHz frequencies, searches for GWs with small correlated interferometers have produced initial bounds \cite{holometer}, there are no established methods to systematically probe the higher frequency part of the GW spectrum, where a variety of interesting GW sources could exist. 

In this Letter, we describe a Levitated Sensor Detector (LSD) based on optically levitated multi-layered dielectric microstructures. The technique can search for  high frequency GWs in the band of $\sim10$-$300$ kHz, extending the frequency reach of existing instruments by over an order of magnitude. Unlike the laser interferometer GW observatories which are limited at high frequency by photon shot noise and typically operate at frequencies below 10 kHz, our approach is limited at high frequency by thermal noise in the motion of the levitated particles and heating due to light scattering. The different frequency scaling of this noise makes the LSD competitive at high-frequencies: while the LIGO sensitivity decreases at higher frequency, the LSD sensitivity improves, enabling a substantial advance by a compact detector \cite{Arvanitaki:2013}.

Optically levitated sensors for high-frequency GW detection were proposed in Ref. \cite{Arvanitaki:2013}.  In this {\em{Letter}} we propose an extension particularly suited for GW detection: using a stack of thin-layered dielectric discs. Stacked disks address a major limiting quantum noise source of the levitated sensor technology --- photon recoil heating --- while at the same time increasing the mass of the levitated object, further increasing sensitivity. Photon recoil heating \cite{GordonAshkin:1980}, recently observed in optical levitation experiments \cite{Jain:2016}, raises the effective temperature of the levitated object and hence degrades force sensitivity \cite{Geraci:2010}. It has been shown theoretically \cite{Arvanitaki:2013,Chang:2012} that if a disc is levitated instead of a sphere, the heating rate can be lowered. The stacked disk approach could result in orders of magnitude of sensitivity improvements, depending on the shape and size of the levitated object. 

The high-frequency GW regime is particularly well-suited for beyond the standard model (SM) physics searches as astrophysical GW sources do not extend above several kHz in frequency due to the physical sizes of stellar-mass compact objects.   A unique high-frequency GW signal can be sourced by macroscopic bound states of axions around light astrophysical BHs \cite{Arvanitaki:2009fg,Arvanitaki:2010sy}.  The QCD axion is a well-motivated beyond-the-SM particle which naturally solves the strong CP problem \cite{Peccei+1977, Weinberg1978, Wilczek1978} and is a dark matter candidate \cite{Preskill:1982cy,Abbott:1982af,Dine:1982ah}. If an ultralight boson, such as the axion, has Compton wavelength of order the BH size, it is produced in exponentially large numbers through superradiance, forming  a ``gravitational atom''.  The axions source coherent, monochromatic GW radiation \cite{Arvanitaki:2010sy,Arvanitaki:2014wva}. The theoretically well-motivated Grand Unified Theory (GUT)-scale QCD axion could form gravitational atoms around stellar-mass BHs 
and source $\sim 100$ kHz GWs, in the optimal sensitivity range of the LSD detector.

GWs could also open a window on dark matter and the early universe. Dark matter  ~\cite{Feng2010,Bertone2005,Profumo2017-Book} is a strong indicator for new physics. Potential candidates include primordial black holes (PBHs), which are primarily searched for through their gravitational signatures \cite{LIGOsubsolarSearch,PBH150914,Sasaki:2018dmp}. The inspiral, merger, or ringdown of sub-solar PBHs will emit GWs in the frequency range accessible by LSD. While the PBH mass spectrum is constrained from existing experiments like supernova lensing \cite{Supernovalensing}, EROS \cite{EROSII}, and MACHO \cite{MACHO}, GW searches in the 10 kHz band provide an independent probe. 

Other predicted sources of high frequency GWs include cosmological sources 
such as axion inflation \cite{Barnaby:2010vf}, cosmic strings \cite{Planck2013},  axionic preheating \cite{Figueroa:2017vfa,Caprini:2018mtu}, and phase transitions \cite{Kamionkowski:1993fg,GarciaGarcia:2016xgv}, as well as other dark matter candidates \cite{Guo:2019ker}. Furthermore, a high-frequency GW observatory based on a network of levitated sensors would yield a high-frequency-GW sky map of our universe.

\begin{figure}[!t]
\begin{center}
\includegraphics[width=0.85\columnwidth]{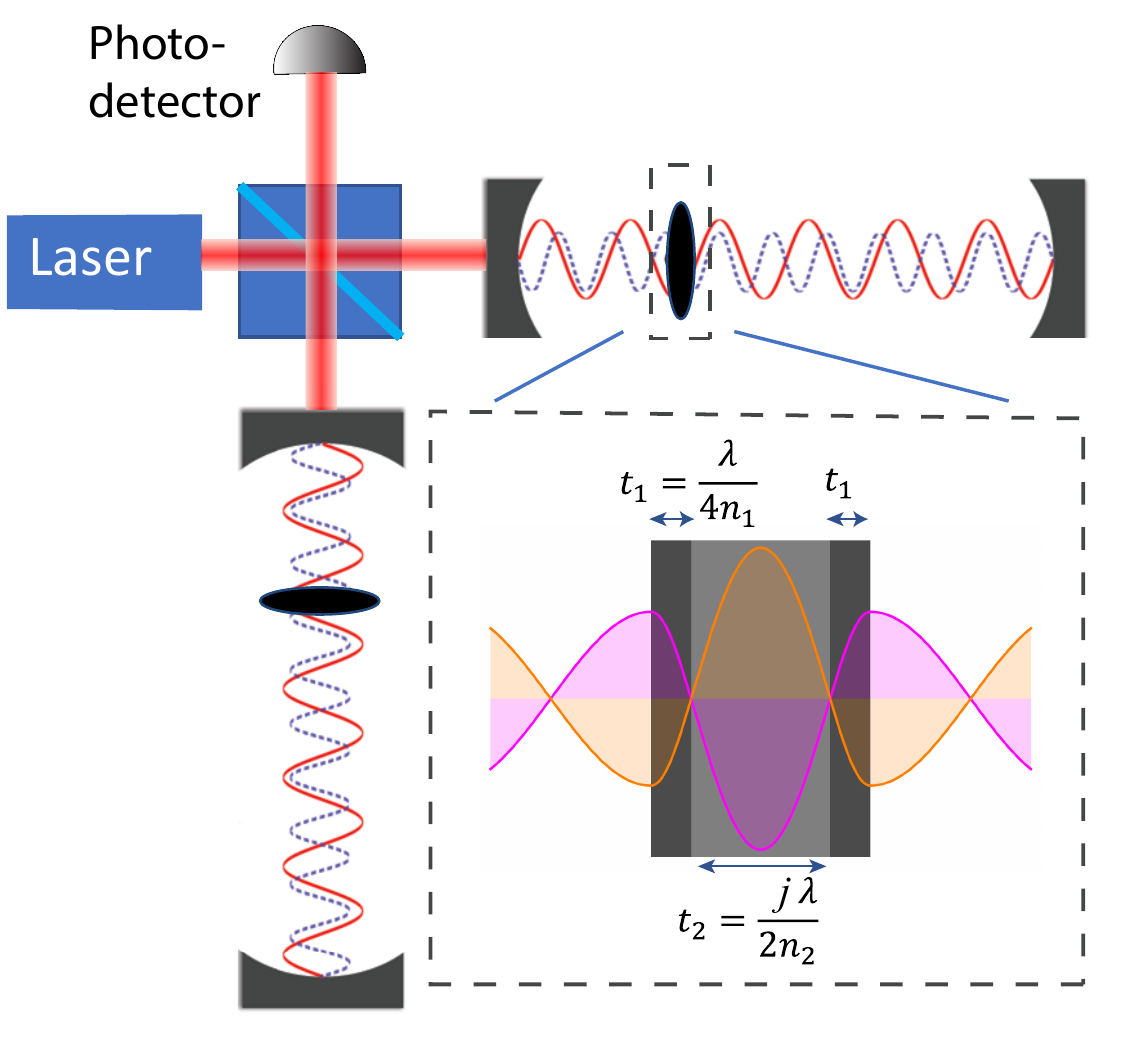}
\caption{Schematic of the levitated sensor detector (LSD) for GW detection at high frequencies. A stack of dielectric discs is optically confined in each Fabry-P\'erot arm of a Michelson interferometer. A secondary beam (dotted-line, not shown in inset) is used to cool and read out the motion of each stack along its respective cavity axis. \textbf{Inset:} Electric field profile of the trapping light as it propagates through the dielectric stack structure supported in each arm of the interferometer. The stack has high-index ($n_1$) end caps and a low-index ($n_2$) spacer with thicknesses $t_1$ and $t_2$, respectively. 
$\lambda$ is the laser wavelength and $j$ is an integer.
\label{setup}}
\end{center}
\end{figure}

\emph{Experimental Setup and Sensitivity---}
We consider a compact Michelson interferometer 
configuration with Fabry-P\'erot arms as shown in Fig. \ref{setup}. A dielectric object is suspended at an anti-node of the standing wave inside each Fabry-P\'erot arm. A second laser can be used to read out the position of the object as well as cool it along the cavity axes, as described for a similar setup in Ref. \cite{Arvanitaki:2013}. The optical potential for this trap is 
$
U=\frac{1}{c}\int{I(\vec{r})(\epsilon(\vec{r})-1)d^3\vec{r}}
$
where \(I\) is the laser intensity, \(\epsilon\) is the relative dielectric constant, and the integration is performed over the extent of the dielectric particle. The trapping frequency along the axis of the cavity is determined by $\omega_0^2=\frac{1}{M}\frac{d^2U}{dx^2} {|}_{x=x_s}$ for a sensor of mass $M$ trapped at equilibrium position $x_s$.  

A passing GW with frequency $\Omega_{\text{gw}}$ imparts a force on the trapped particle \cite{Arvanitaki:2013}, and 
when  \(\omega_0\) $=\Omega_{\text{gw}}$, it will be resonantly excited. Unlike a resonant-bar detector, \(\omega_0\) is widely tunable with laser intensity. 
The second cavity arm permits rejection of common mode noise, for example from technical noise in the laser or vibration.

The minimum detectable strain $h_{\rm{limit}}$ for a particle with center-of-mass temperature $T_\mathrm{CM}$ is approximately \cite{Arvanitaki:2013}
\beq
h_{\mathrm{limit}}=\frac{4}{\omega_0^2L}\sqrt{\frac{k_BT_{\mathrm{CM}}\gamma_gb}{M}\left[1+\frac{\gamma_{\mathrm{sc}}}{N_i\gamma_g}\right]}H\left(\omega_0\right), \label{eq:noise}
\eeq
where the cavity response function $H\left(\omega\right)=\sqrt{1+\left(2\mathcal{F}/\pi\right)^2\sin^2{\left(\omega L/c\right)}}$ 
for a cavity of length $L$ and finesse \(\mathcal{F}\). Here $N_i=k_BT_\mathrm{CM}/\hbar\omega_0$ is the mean initial phonon occupation number of the center-of-mass motion, where $\hbar$ is Planck's constant and $k_B$ is Boltzmann's constant. $\gamma_g = \frac{32P}{\pi \bar{v} \rho t}$ is the gas damping rate at pressure $P$ with mean gas speed $\bar{v}$ for a disc of thickness $t$ and density $\rho$, and $b$ is the bandwidth.

The photon recoil heating rate \cite{Arvanitaki:2013,Chang:2012} 
$\gamma_{sc}=\frac{V_c\lambda\omega_0}{4L}\frac{1}{\int{dV(\epsilon-1)}}\frac{1}{\mathcal{F}_{\rm{disc}}}
$ is inversely proportional to the disc-limited finesse $\mathcal{F}_{{\rm{disc}}}$, i.e. \(2\pi\) divided by the fraction of photons scattered by the disc outside the cavity mode. The integral is performed over the extent of the suspended particle. Here $V_c$ is the cavity mode volume \cite{Arvanitaki:2013}. For a nanosphere, the elastic scattering is like that of a point dipole, and nearly isotropic, resulting in a random recoil which leads to a momentum diffusion process for the center of mass of the sphere \cite{Jain:2016}. For a disc, if the beam size is smaller than the radius of the object and the wavefront curvature at the surface is small, the scattered photons acquire a stronger directional dependence and tend to be recaptured into the cavity mode. 
This reduces the variance of the recoil direction of the levitated object caused by the scattered photons.

Both of the damping rates that contribute to sensitivity in Eq. \ref{eq:noise} scale inversely with the thickness of the levitated disc, for thickness smaller than radius. 
In the gas-dominated regime, $\gamma_{\rm{sc}} \ll N_i\gamma_g$, the sensitivity scales as $\sqrt{1/Mt}$ at fixed frequency. 
For sufficiently low vacuum, the sensitivity becomes photon-recoil-limited, and 
the strain sensitivity goes as \(1/M\sqrt{\mathcal{F_\mathrm{disc}}}\).\footnote{For these scalings we have assumed a similar density throughout the levitated object.}

We demonstrate that it is possible to increase the mass of the levitated object, and hence the sensitivity to GWs, without substantially increasing the photon recoil rate by using a \textit{stacked} disc geometry. The thickness of each layer can be chosen to attain perfect transmission, and the high-index sections serve as ``handles'' since they have a stronger affinity to the antinodes of an optical standing wave. Multiple reflections within the stack further enhance the optical trapping potential.

As a proof-of-principle, we consider a dielectric stack in a 3-layer configuration with high-index endcaps of thickness $t_1=\lambda/4n_{1}$ on a low-index spacer cylinder of length $j\lambda/2n_2$, where $n_{1}$ and $n_{2}$ are the index of refraction of the endcaps and spacer, respectively, and $j$ is an integer. Specifically, we chose a dielectric stack consisting of a silica ($n_2=1.45$) spacer with Si ($n_1=3.44$) endcaps. The entire stack is of radius $75$ $\mu$m and the beam radius is chosen to be $37.5$ $\mu$m at the location of the stack. Proposed experimental parameters are shown in Table \ref{table2}.  

\begin{table}[!t]
\begin{center}
  \begin{tabular}{@{}cccc@{}}
  \hline
  \hline
  Parameter & Units & \hspace{5pt} $\omega_0/2\pi=10$ kHz \hspace{5pt}   & $\omega_0/2\pi=100$ kHz \\
  \hline
$\lambda$ & $\mu$m & $1.5$ & $1.5$ \\
$I_0$ & W/$m^2$ & $2.2 \times 10^8$ & $2.2 \times 10^{10} $\\
$N_i \gamma_g $ & Hz & $1.7$ & $0.17$\\
$ \gamma_{sc} $ & Hz & $0.005$ & $0.05$ \\
$h_{\rm{min}}$ & 1$/\sqrt{\rm{Hz}}$ & $7.6 \times 10^{-21}$ & $1.02 \times 10^{-22} $ \\
  \hline
  \hline
  \end{tabular}
\caption{\label{table2} Experimental parameters for trapping of a $75$ $\mu$m radius stack with $14.58$ $\mu$m thick silica spacer (corresponding to $j=28$) and quarter-wave $110$ nm thick Si endcaps in a cavity of length $L=10$ m at $P=10^{-11}$ Torr and room temperature. $I_0$ is the peak laser intensity striking the disc and $h_{\rm{min}}=h_{\rm{limit}}/\sqrt{b}$ is the strain sensitivity where $b$ is the measurement bandwidth.}\end{center}
\end{table}

To estimate $\mathcal{F}_{\rm{disc}}$ for the dielectric multilayer stack geometry we employ a finite element Greens Dyadic method based on the pyGDM2 toolkit \cite{pyGDM2}.
We compute the 3D scattering behavior of dielectric stacks for a variety of geometries. As a benchmark, we simulate SiO$_2$ discs and nanospheres of numerically-tractable size and find them to agree with analytical limits. 
To determine $\mathcal{F}_{\rm{disc}}$, we assume that the photons which scatter into twice the $1/e^2$ beam radius at the cavity end mirror are recaptured in the cavity mode, justified for the stack and beam radii considered here. 
\begin{figure}[!t]
\centering

\includegraphics[width=0.45\textwidth,height=0.45\textheight,keepaspectratio]{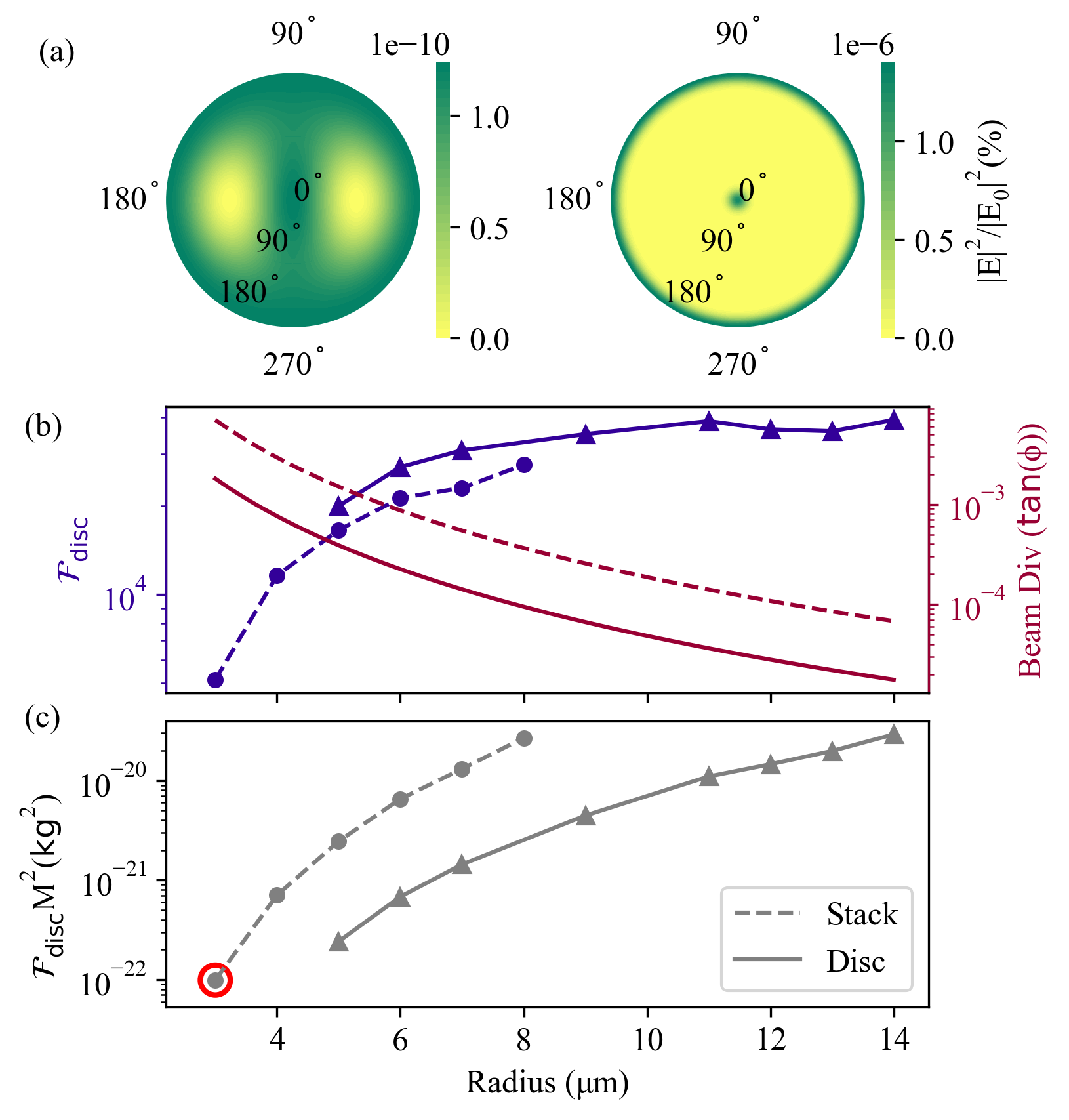}

\caption{(a) (left) Far-field scattered light intensity distribution for a nanoparticle which acts as a point-like Rayleigh scatterer; and (right) for a dielectric Si/SiO$_2$/Si stack with $j=1$ and radius 3 $\mu$m, where the laser beam waist is chosen to be one half the stack radius.  (b) Disc-limited finesse $\mathcal{F}_{{\rm{disc}}}$ and beam divergence angle $\phi$ at the object surface for Si discs (solid line) and Si/SiO$_2$/Si stacks  with $j=1$ (dashed line) for varying radii. (c) $\mathcal{F}_{{\rm{disc}}} \times M^2$ (figure of merit in the photon-recoil-dominated regime
) vs. radius. 
The red circled point corresponds to the stack considered in (a).
\label{fig:finesse}}

\end{figure}

\begin{figure}[!t]
\begin{center}
\includegraphics[width=1.0\columnwidth]{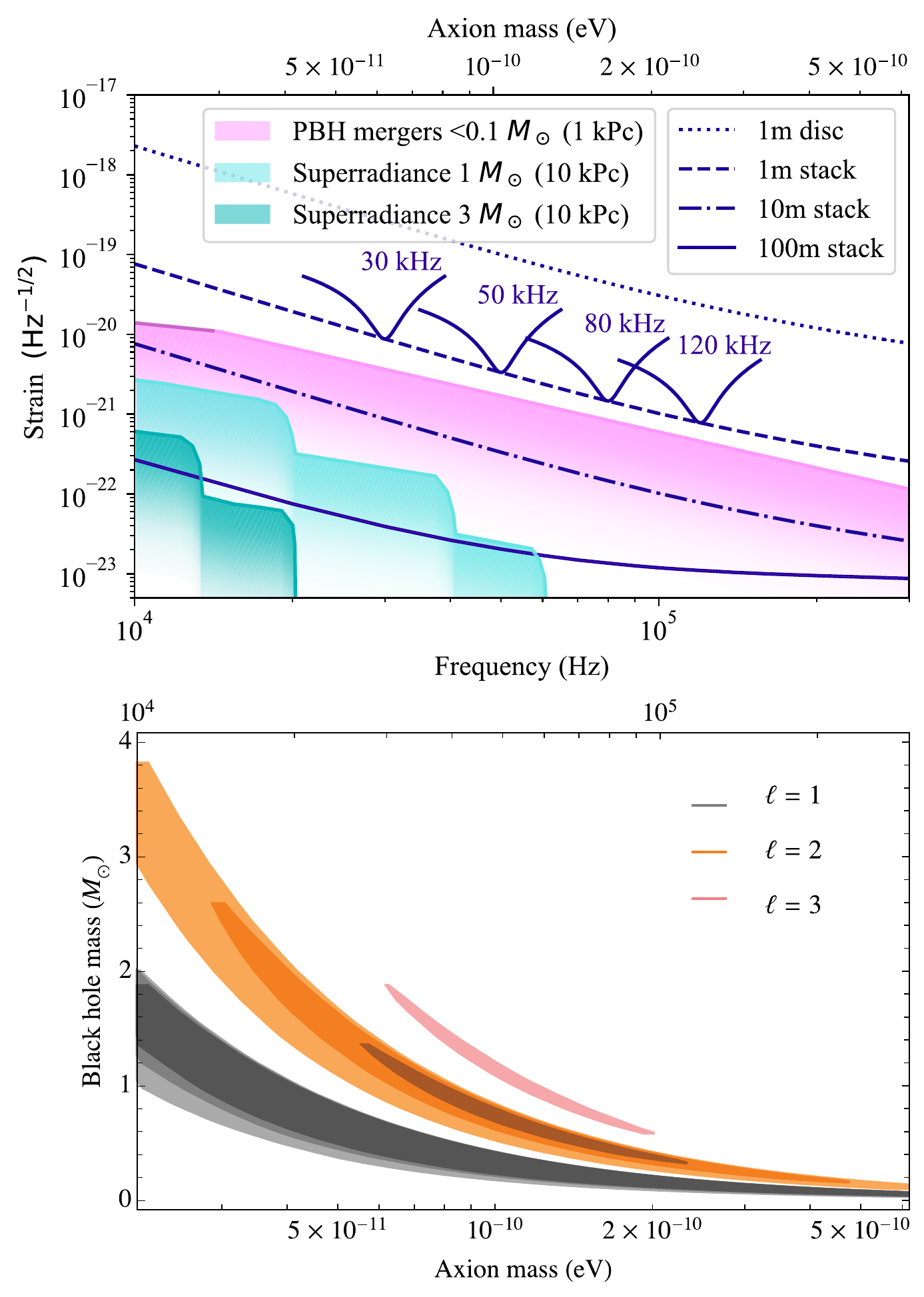}
\caption{(upper) Strain sensitivity for optically-levitated microdiscs (dotted) or stacked discs (dashed), at design sensitivity for the 1-m prototype instrument. The sensitivity curves are formed as the locus of the minima of the sensitivity from a single realization of the tunable optical trap frequency (examples shown for illustration at 30, 50, 80, and 120 kHz). The cyan shaded regions denote predicted signals due to GWs produced from axions around BHs in our galaxy within $10$ kpc for $10^6$s coherent integration time. The pink area shows the expected strain from inspiraling and merging PBHs at distances $\geq1$ kpc. Also shown is projected sensitivity for a future 10-m room-temperature (dash-dot) and 100-m cryogenic setup (solid). (lower) Reach to axion annihilation signals of the 100-m stack LSD setup. The shaded regions indicate where the reach exceeds 10 (light), 30 (medium), 50 (dark) kpc for a BH with initial spin $a_* = 0.9$ as a function of axion and BH mass. The three bands correspond to the $\ell = m = \{1, 2, 3\}$ SR levels. The $\ell=3$ level exceeds only 10 kpc.
\label{strain}}
\end{center}
\end{figure}

\emph{Results---}
We show the results of the scattering simulations in Fig. \ref{fig:finesse}. In Fig. \ref{fig:finesse}(a) we show the distribution of scattered light in the far-field for a nanoparticle which acts as a point Rayleigh scatterer as well as for a dielectric stack of radius 3 $\mu$m. The laser beam waist radius is chosen to be one half the stack radius. Taking this scattering into account allows us to compute the disc-limited finesse. In Fig. \ref{fig:finesse}(b), we show the disc-limited finesse $\mathcal{F}_{{\rm{disc}}}$ and beam divergence at the object surface for Si discs and Si/SiO2 stacks for structures of varying radii. As expected,  $\mathcal{F}_{{\rm{disc}}}$ increases as the beam divergence decreases.  
For our current setup of a $75$ $\mu$m stack, we conservatively estimate $\mathcal{F}_{\rm{disc}}$ as $4\times 10^4$, the value calculated for a $14$ $\mu$m radius disc. The $\mathcal{F}_{\rm{disc}}$ calculation for larger radii is limited by computational memory, but our current results at smaller radii up to $14$ $\mu$m indicate an increasing trend (see Fig. \ref{fig:finesse}b).
The stack $\mathcal{F}_{\rm{disc}}$ is large enough such that for our parameters, we stay in the gas-damping-limited regime, where the sensitivity is independent of $\mathcal{F}_{\rm{disc}}$ and improves with both mass and thickness. In the photon-recoil-limited regime, the figure of merit $\mathcal{F}_{{\rm{disc}}} \times M^2$ is shown in Fig \ref{fig:finesse}(c). The better performance from using a stack comes from having a larger mass with a relatively small reduction in $\mathcal{F}_{\rm{disc}}$.

In Fig. \ref{strain} we show the estimated reach in strain sensitivity for the setup shown in Table \ref{table2}. The $300$ kHz upper limit is chosen due to expected limitations from absorbed laser power by the suspended particle. In practice we estimate that the stack thicknesses need to be precise at the $\sim$ 1.5 nm and 0.5 nm level to ensure $>99$ \% and $99.9$ \% transmission, respectively.  We assume vacuum of $10^{-11}$ Torr and room temperature for all cases except we assume cryogenic ($4$ K) for an optimized $100-$m facility.  For our parameters which yield minimal recoil-heating, the sensitivity remains in the gas-damping-dominated regime 
despite the relatively large mass of the the levitated particle.
Since LSD is a resonant detector, we show the strain sensitivity in Fig. \ref{strain} as the locus of best sensitivity for each tuned configuration. The resonant width (i.e. detector $Q$) is tunable via laser cooling as discussed in Ref. \cite{Arvanitaki:2013}, and $h_{\rm{limit}}$ is $Q-$independent given sufficient displacement sensitivity \cite{Arvanitaki:2013}. $Q=10$ is shown for illustration.  We show also the predicted signals from BH superradiance and PBH inspirals and mergers.

\emph{Gravitational waves from PBH mergers---} If sub-solar blackholes are observed to be merging, they are likely to be primordial in origin, forming part of the galactic dark matter. The pink area in Fig. \ref{strain} shows the expected GW strain from inspiraling and merging PBHs at a distance of 1 kpc. The dark pink line shows the strain from the inspiral of two \(0.1 M_\odot\) BHs and terminates at $\sim 14.4$~kHz, the GW frequency corresponding to the innermost stable circular orbit (ISCO) of the binary. Binaries of lighter BHs merge at higher frequencies, and the locus of their ISCO frequencies forms the boundary of the possible PBH signal space, shown in pink. Weaker signals from earlier inspiral stages, farther source distances and suboptimal source orientations form the shaded area.

\emph{Gravitational Waves from Axion Superradiance---} The angular momentum and energy of rotating astrophysical BHs can be converted into gravitationally bound states of exponentially large numbers of ultralight bosons through BH superradiance \cite{Damour:1976kh,Ternov:1978gq,Zouros:1979iw,Detweiler1980,Arvanitaki:2009fg,Arvanitaki:2010sy}, resulting in a unique tool to search for ultralight scalar~\cite{Arvanitaki:2009fg,Arvanitaki:2010sy,Arvanitaki:2014wva,Brito:2014wla,Brito:2015oca,Arvanitaki:2016qwi,Brito:2017wnc,Brito:2017zvb,Baumann:2018vus} and vector \cite{Pani:2012bp,Witek:2012tr,Baryakhtar:2017ngi,East:2017mrj,Siemonsen:2019ebd} particles beyond the SM. The resulting ``gravitational atom" has bound levels characterized by their principal, orbital, and azimuthal quantum numbers $\{n,\ell, m\}$, and by a  ``fine structure constant" $
    \alpha \equiv G M_\text{BH} \mu/( \hbar c^3)\simeq 0.2 \left(\frac{M_\text{BH}}{1 M_\odot}\right)\left(\frac{\mu}{3\times 10^{-11} \text{eV}}\right)$
where  $M_\text{BH}$ is the BH mass and $\mu$ is the axion rest-energy. 

The superradiance (SR) condition is given by
$\alpha~\leq~\frac{m}{2}\frac{a_*}{1+\sqrt{1-a_*^2}}$, where $0 \leq a_*<1$ is the  dimensionless BH spin. Levels which satisfy the SR condition grow exponentially in occupation number once the BH is formed, until order one of the BH angular momentum is extracted and the BH spin saturates the SR condition~\cite{Damour:1976kh,Ternov:1978gq,Zouros:1979iw,Detweiler1980}.  For more details see \cite{Arvanitaki:2010sy,Arvanitaki:2014wva,Arvanitaki:2016qwi,Brito:2015oca} and references therein.

Axions from a single level annihilate, sourcing continuous, monochromatic GWs with angular frequency of approximately twice the axion rest energy $
 f_\text{GW}\simeq \frac{\mu}{\pi\hbar} \simeq145 \text{ kHz} \left(\frac{\mu}{3\times 10^{-10} \text{ eV}} \right)\,
$~\cite{Arvanitaki:2009fg,Arvanitaki:2010sy}.
While searches with LIGO data are underway for bosons with rest energy up to $4\times10^{-12}$~eV \cite{Tsukada:2018mbp,Dergachev:2019oyu,Palomba:2019vxe,Zhu:2020tht}, high-frequency detectors are necessary to observe the annihilation signal from theoretically well-motivated QCD axions with decay constant $f_a$ near the GUT scale, $\mu\simeq 3\times10^{-10}\,\mathrm{eV}~(2\times10^{16}\mathrm{GeV}/f_a)$.

The angle-averaged signal strain from  the fastest-growing $\ell=m=1$ axion cloud level\footnote{At fixed $\ell$, the growth and annihilation rates are maximal for $\ell = m$, which is assumed throughout.} at distance $r$ scales, at leading order in $\alpha$, as \begin{align}
\label{eqn:strain}
h\sim 10^{-24}\left( \frac{\Delta a_*}{0.1}\right)\left( \frac{10 \text{kpc}}{r}\right)\left( \frac{M_\text{BH}}{1 M_\odot}\right) \left( \frac{\alpha}{0.2}\right)^7,
\end{align}
with difference between initial and final spin $\Delta a_*$.

As the cloud depletes through annihilations, the signal decreases over time as $h(t) = \frac{h (t = 0)}{1+t/\tau_{\text{sig}}}\,$
where the typical signal duration $\tau_{\text{sig}}$ ranges from $10^6$~s to $10^{12}$~s for the parameter space under consideration. The signal frequency undergoes a positive drift  $ \dot{f}$ 
which could aid in characterizing a potential signal~\cite{Arvanitaki:2016qwi, Baryakhtar2020}. The drift is  small compared to the detector bandwidth over the integration time,  $ \dot{f}\lesssim 4\times 10^{-8} \text{Hz/s} $.

Figure~\ref{strain} (upper) shows the maximum integrated strain of axion annihilation signals $h\, t_\text{int}^{1/2}$ from a BH within $10$~kpc with initial  spin $a_*^{\mathrm{init}} = 0.9$, assuming a coherent integration time of $t_\text{int}=10^6 $~s. The envelope consists of levels $\ell=1,2,3$, with $\ell=3$ reaching higher axion masses, and BH masses of $1M_{\odot}$ and $3 M_{\odot}$, with weaker signals arising from more distant and heavier BHs.

In Fig.~\ref{strain} (lower) we show the LSD reach for annihilation signals. Heavier axions can only form clouds of a given angular momentum around relatively lighter BHs due to the SR condition. At fixed BH mass,  heavier axions can form clouds only in levels with higher $\ell$. As 
there is thought to be a gap in compact object masses with no BHs of $ M_{\mathrm{BH}} \lesssim 5 M_\odot$ formed \cite{Bailyn:1997xt,Ozel:2010su,kreidberg2012mass,belczynski2012missing} (although see new evidence of mass-gap compact objects \cite{Abbott:2020khf,Thompson637,Margalit:2017dij}), 
 it is particularly interesting to search for signals from $\ell>1$ to reach new, heavier axion parameter space. 

 Signals for higher-$\ell$ levels are weaker, with $h \propto (\alpha/\ell)^{2\ell+5}$ at leading order in $\alpha$, and last longer. In Fig.~\ref{strain}, we use the  fully numerical annihilation power calculations of Ref. \cite{Yoshino2014}. We
do not take into account axion self-interactions \cite{Arvanitaki:2010sy,Gruzinov2016,Baryakhtar2020}, which may affect the dynamics of higher levels and need to be included in a complete search analysis \cite{Baryakhtar2020}.

\emph{Discussion ---}
Current GW observatories such as Advanced LIGO do not search for GWs over 10 kHz.  Our approach enables a search for well-motivated beyond the standard model sources of GWs such as the GUT-scale QCD axion, which could naturally exist at these frequencies.
Looking forward, the few kHz frequency band is the prime region for GW emission from the post-merger dynamics of the compact object resulting from a binary neutron star inspiral \cite{NSEoS1,NSEoS2}. 
Using even larger levitated masses could lead to further sensitivity improvements, enabling deeper exploration of physics such as the neutron star equation of state. The approach described in this \emph{Letter} will have a vast discovery potential in uncharted GW frequency parameter space. 

\emph{Acknowledgements}
We would like to thank A. Arvanitaki and P. Barker for useful discussions. MB is supported by the James Arthur Postdoctoral Fellowship. MT is partially supported by the Stanford Physics Department Fellowship. AG is supported in part by NSF grants PHY-1806686 and PHY-1806671, the Heising-Simons Foundation, the John Templeton Foundation, and ONR Grant N00014-18-1-2370. AG, SL, and VK are supported by the W.M. Keck Foundation. This work used the Extreme Science and Engineering Discovery Environment (XSEDE) 
at the Pittsburgh Supercomputing Center through allocation TG-PHY190038, and the Quest computing facility at Northwestern.

\emph{Author contributions}
NA and GW contributed to the optical trapping calculations. MT, MB, and NA estimated sources. AG, SL, VK supervised the project.
All authors contributed to  discussions and writing.

\bibliography{library-apd,library_jw,library_geraci,PBH_bib,BHSR_bib} 

\end{document}